# Investigation of Negative Capacitance Vertical Nanowire FETs Based on SPICE Model at Device-Circuit Level

Weixing Huang, Huilong Zhu, Kunpeng Jia, Zhenhua Wu, Xiaogen Yin, Qiang Huo and Yongkui Zhang.

*Abstract*—In this study, a SPICE model for negative capacitance vertical nanowire field-effect-transistor (NC VNW-FET) based on BSIM-CMG model and Landau-Khalatnikov (LK) equation was presented. Suffering from the limitation of short gate length there is lack of controllable and integrative structures for high performance NC VNW-FETs. A new kind of structure was proposed for NC VNW-FETs at sub-3nm node. Moreover, in order to understand and improve NC VNW-FETs, the S-shaped polarization-voltage curve (S-curve) was divided into four regions and some new design rules were proposed. By using the SPICE model, device-circuit co-optimization was implemented. The co-design of gate work function (WF) and NC was investigated. A ring oscillator was simulated to analyze the circuit energy-delay, and it shown that significant energy reduction, up to 88%, at iso-delay for NC VNW-FETs at low supply voltage can be achieved. This study gives a credible method to analysis the performance of NC based devices and circuits and reveals the potential of NC VNW-FETs in low-power applications.

## I. INTRODUCTION

THE negative capacitance field-effect-transistor (NCFET) can overcome the so-called Boltzmann tyranny which defines the fundamental thermionic limit of the sub-threshold slop (SS) at 60 mV/decade at 300K. Consequently, NCFET can break through the lowest limit of SS to lower the supply voltage and overall power consumption [1]-[4]. The vertical nanowire field-effect-transistor (VNW-FET), with great short channel effects suppression, has potential to reduce power consumption and increase the integration density of integrated circuits (ICs). Benefit from those advantages, NC VNW-FETs get a lot of attentions on application for sub-3nm node. Moreover, due to enormous complexity of NCFETs, SPICE model and good understanding for NCFETs are imperative to implement circuit

This work is supported by the Academy of Integrated Circuit Innovation under grant No Y7YC01X001.

The authors are with the Key Laboratory of Microelectronics Device & Integrated Technology, Institute of Microelectronics, Chinese Academy of Sciences, Beijing 100029, China, and with the University of Chinese Academy of Sciences, Beijing 100049, China. (e-mail: jiakunpeng@ime.ac.cn,wuzhenhua@ime.ac.cn,zhuhuilong@ime.ac.cn) .

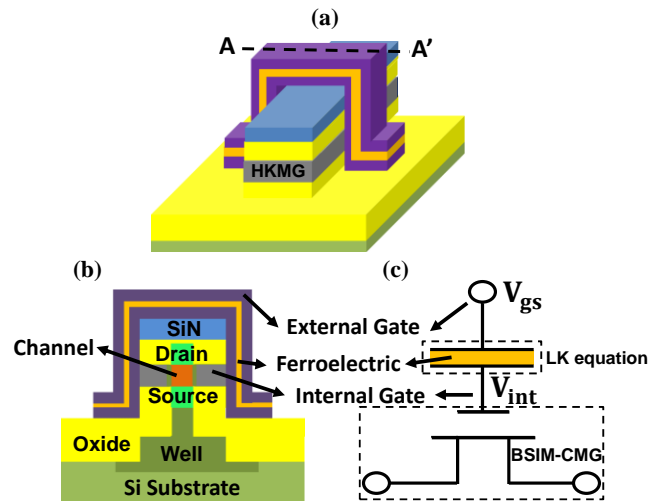

Fig.1 NC VNW-FET device structure. (a) The 3D structure of NC VNW-FET (there are three parallel nanowires.). (b) Cross section across AA' direction. (c) Equivalent circuit diagram of NC VNW-FET.

design, it is above rubies to investigate and build SPICE model for NCFETs to analysis the performance of NCFET-based circuit.

Recently, device modeling and small-scale circuit analysis for NCFETs have been performed [5]-[9]. However, there is a lack of the characteristics analysis of ultra-size device which is applied for sub-3nm node such as NC VNW-FETs. In this study, a physics-based SPICE model for NC VNW-FET based on BSIM-CMG model and Landau-Khalatnikov (LK) equation was presented. The industry standard BSIM-CMG model accurately models the second-order effects of aggressively scaled devices and accounts for short-channel effects [10]. An accurate conventional VNW-FET model is very important to develop a model for NC VNW-FETs because the LK equation depends on the charge of the conventional VNW-FETs. Time-dependent LK equation was used in the model to obtain the relationship of polarization-voltage (P-V) of the ferroelectric and the ferroelectric parameters which were used in LK equation were calibrated from our experimental $Hf_{0.5}Zr_{0.5}O_2$ data. The LK equation is coupled with a BSIM-CMG model, whose parameters were extracted from the TCAD simulated conventional VNW-FETs, in a self-consistent form to predict the characteristics of NC VNW-FET.

Moreover, based on International Roadmap for Device and Systems (IRDS) 2017 [11], the physical gate length of VNW-FETs for high performance logic application will be scaled to 12nm. So, there is no sufficient space for ferroelectric layer owing to such short gate length. Although ultrathin ferroelectric film can be integrated, but ultrathin ferroelectric film is not easy to show the NC characteristics, and is easy to cause reliability problem, such as serious gate dielectric leakage current. Suffering from the limitation of gate length of VNW-FETs there is lack of controllable and integrative structures to combine VNW-FETs with ferroelectric material. In this study, a new kind of structure was proposed for high performance VNW-FETs integrated with ferroelectric film at sub-3nm node. Based on this kind of structure the area of NC can be adjusted for capacitance matching easily. Of note, the structure, proposed in this study, can be fabricated by using our in-house process [12].

The contributions of this study can be summarized as follows.
1) A physics-based SPICE model for NC VNW-FETs at sub-3nm node based on BSIM-CMG and LK equation was built.
2) A new kind of structure was proposed for high performance NC VNW-FETs at sub-3nm node.
3) The S-shaped polarization-voltage curve (S-curve) was divided into four regions and explained from mathematical model and physical mechanism. Based on S-curve four regions theory, a few design rules of NCFETs were proposed.
4) Based on DC analysis of NC VNW-FET inverter, the co-design of gate work function (WF) and NC was investigated.
5) A ring oscillator was simulated to analyze the circuit energy-delay, and it shown that signification energy reduction, up to 88%, at iso-delay for NC VNW-FETs can be achieved.

## II. NC VNW-FET SPICE MODEL

The simulated device structure of an NC VNW-FET and equivalent circuit diagram are shown in Fig.1. Based on the showed structure in Fig.1 (a), the value of NC ($C_{FE}$) can be adjusted by changing area of NC ($A_{FE}$) for capacitance matching easily. As shown in Fig.1 (c), NC VNW-FET can be seen as a series-wound combination of VNW-FET with NC, and the NC VNW-FET SPICE model was built based on this concept.

The conventional VNW-FETs were simulated by TCAD firstly, which were used to extract parameters of BSIM-CMG model. In this study, based on IRDS 2017, the physical gate length and channel diameter of VNW-FETs was set to 12nm and 6nm respectively at sub-3nm node. The doped silicon (As/1e20cm$^{-3}$ for nVNW-FET and B/1e20cm$^{-3}$ for pVNW-FET) was used as source/drain. The source and drain doping slops were generated by rapid thermal annealing (RTA) at 1100°C. The undoped silicon was used as channel material. The gate oxide thickness (EOT) and overlay length is 0.8nm and 2nm respectively. Moreover, the contact and source/drain epi-diffused resistance is set to 4000Ω. In aggressively scaled devices of sub-3nm node, prominent quantum confinement makes the typical bulk material characteristics, such as bandgap, effective mass and carrier mobility, et al., change significantly depending on the shape and size of the channel. Conventional TCAD models, based on bulk band structure and drift-diffusion transport equation, are inadequate to capture the essential device characteristics unless re-calibrating the quantum correction model, piezo-resistivity model, liw-field and high-field mobility models, et al., for every change of the device design, which is quiet impractical. In this study, the sub-band

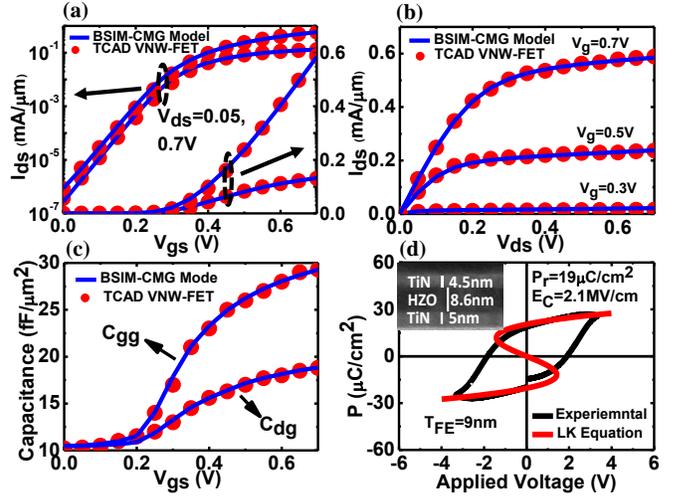

Fig.2 conventional VNW-FET model validation: agreement between TCAD VNW-FET and BSIM-CMG model. (a) $I_{ds}$-$V_{gs}$ characteristics of the nVNW-FET; (b) $I_{ds}$-$V_{ds}$ characteristics of the nVNW-FET; (c) C-V characteristics of the nVNW-FET. (d) P-V curve and cross-sectional TEM image of $H_{0.5}Z_{0.5}O_2$ (embedded figure).

Boltzmann transport equation which is coupled to the Schrö$d$inger-Poisson solver for quantum confinement effect was performed to consider the quantum effects in nano-scale transistors. Furthermore, the phonon scattering and the surface roughness scattering were taken into consideration. An excellent model agreement between BSIM-CMG model and TCAD simulation for conventional VNW-FETs is shown in Fig.2 (a)-(c).

To build the SPICE model for NC VNW-FETs, time-dependent LK equation was introduced to describe the proprieties of ferroelectric films in transistors, given as follows:

$$E = 2\alpha P + 4\beta P^3 + 6\gamma P^5 + \rho \frac{dP}{dt} \quad (1)$$

here $\alpha$, $\beta$ and $\gamma$ are static coefficients and $\rho$ is the kinetic coefficient. Let $V_{FE}$, $A_{FE}$ and $T_{FE}$ be the voltage across ferroelectric, ferroelectric area and thickness respectively. Combining equation (1) with $E = V_{FE}/T_{FE}$ and $Q_{FE} = A_{FE}P$,:

$$V_{FE} = T_{FE}(2\alpha \frac{Q_{FE}}{A_{FE}} + 4\beta \frac{Q_{FE}^3}{A_{FE}^3} + 6\gamma \frac{Q_{FE}^5}{A_{FE}^5}) + \rho \frac{T_{FE}}{A_{FE}} \frac{dQ_{FE}}{dt} \quad (2)$$

The core of the SPICE model is the coupling of BSIM-CMG model and LK equation that is based on the charge balance principle. In NC VNW-FET, it means that the charge on the internal gate ($Q_{int}$) is equal to that on the ferroelectric layer ($Q_{FE}$). So, with a given $Q_{int}$ the $V_{FE}$ can be calculated by equation (2). The high resolution transmission electron microscopy (HRTEM) image of TiN/$Hf_{0.5}Zr_{0.5}O_2$/TiN capacitor stack structure is shown in the insert of Fig.2 (d). The P-V curve and the extracted "S-curve" of the ferroelectric layer were plotted in Fig.2 (d). Then, with the ferroelectric parameters extracted from our experimental data the internal gate voltage ($V_{int}$) can be acquired by equation $V_{int} = V_{gs}$-$V_{FE}$ ($V_{gs}$ is external gate voltage as shown in Fig.1 (c)). Thus, the electrical characteristics of NC VNW-FETs can be acquired by applying $V_{int}$ on conventional VNW-FETs.

## III. DEVICE ANALYSIS

To explain the behavior of NC VNW-FET-based logic circuits, the device characteristics and their dependence on $C_{FE}$

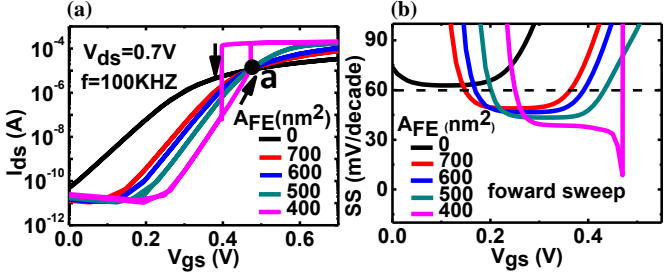

Fig.3 The characteristics of NC VNW-FETs. (a) The $I_{ds}$-$V_{gs}$ characteristics of NC VNW-FETs for different $A_{FE}$. (b) The SS characteristics of NC VNW-FETs for different $A_{FE}$.

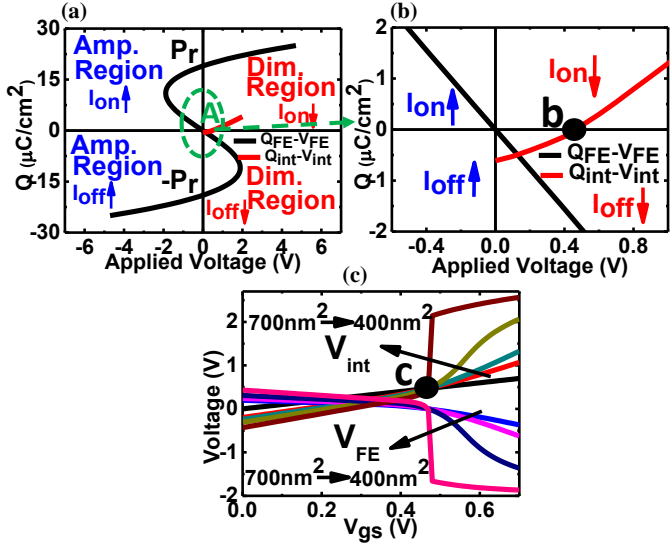

Fig.4 Four regions of "S-curve". (a) Four regions of "S-curve" and $Q_{int}$-$V_{int}$ curve. (b) Zoomed portion of region A in (a). (c) Different potential components of NC VNW-FETs: $V_{gs}$, $V_{FE}$, and $V_{int}$.

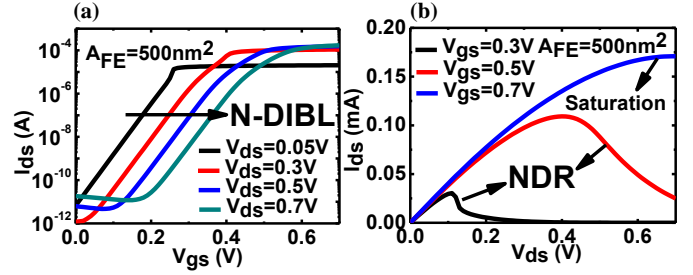

Fig.5 (a) Negative DIBL effect at $A_{FE}$ = 500nm$^2$. (b) NDR effect at $A_{FE}$ = 500nm$^2$ for low $V_{gs}$.

are discussed in this section. To ensure the credibility of data, the $A_{FE}$ was selected as the main variable of $C_{FE}$, owing to the properties of ferroelectric layer is sensitive to the layer thickness and component. The transient characteristics of NC VNW-FETs with different $A_{FE}$ ($A_{FE}$ = 0nm$^2$ represents for conventional VNW-FET) are shown in Fig.3. It is evident that as $A_{FE}$ decreases, switching characteristics become steeper due to the amplification in $V_{int}$ provided by NC [13]-[14]. However, the NC VNW-FET is unstable when $A_{FE}$ is smaller than a certain critical area. The SS characteristics of NC VNW-FET and conventional VNW-FET are shown in Fig.3 (b). The minimum SS has reduced to 43mV/decade with negligible hysteresis at $A_{FE}$ = 500nm$^2$. Moreover, comparing to conventional VNW-FET the $V_t$ of NC VNWFET is increasing with the decrease of $A_{FE}$ at same gate WF as shown in Fig.3 (a). It is worth noting that all $I_{ds}$-$V_{gs}$ curves intersect at the same point named "a" or representing "attractor" in this study and the value of $V_{gs}$ at point "a" is $V_a$. This phenomenon was also found in experiments [15]-[16].

To explain the crossover point of $I_{ds}$-$V_{gs}$ curves, the S-curve was divided into four regions as shown in Fig.4 (a), and each region represents a specific working mode. For NCFET, the equation $V_{gs} = V_{FE} + V_{int}$ is a basic assumption. When the transistor works in the first or fourth quadrant (the second and third quadrant) of the S-curve, the $V_{FE}$ is a positive (negative) value. Thus, comparing to conventional VNW-FET the $V_{int}$ decreases (increases) at same $V_{gs}$. Thus the first and fourth quadrant (the second and third quadrant) are named as "Diminution Region" ("Amplification Region"). Further analysis indicates that the behavior of the on current ($I_{on}$) and the off current ($I_{off}$) for NC VNW-FET are different in the four regions because of amplification and diminution of $V_{int}$ in different region. For example, the decrease (increase) of $V_{int}$ will induce the decrease (increase) of $I_{off}$ in subthreshold region, but decrease (increase) of $I_{on}$ in strong inversion region as shown in Fig.4 (a). Physically, the point "a" is corresponding to the unstable point at which the NC polarization starts to change sign and result in increasing or decreasing of the current ($I_{ds}$) for different NC VNW-FETs relative to that of the conventional VNW-FET. Therefore, it is desirable that the threshold voltage $V_t$, is set as close as possible to $V_a$ to obtain small SS.

In the simulation, the $Q_{int}$ is equal to the $Q_{FE}$ based on charge balance principle. From Fig.4 (b) when $V_{gs} < V_a$ (at point "b", actually, the values of $V_{gs}$ are equivalent at point "a", "b", and "c") the $Q_{int} = Q_{FE}$ is located in the fourth quadrant, thus $V_{FE}$ is a positive value, and $V_{int}$ decreases which are demonstrated in Fig.4 (c). So, the $I_{ds}$ decreases due to the decrease of $V_{int}$ which are shown in Fig.3 (a). Oppositely, when $V_{gs} > V_a$ the $Q_{int} = Q_{FE}$ is located in the second quadrant, thus $V_{FE}$ is negative, and $V_{int}$ increases. Similarly, the $I_{ds}$ increases due to the increase of $V_{int}$ which are also shown in Fig.3 (a). Based on the above analysis, the $Q_{int} = Q_{FE}$ should be hold in the second and fourth quadrant of the S-curve for high performance NCFET. If the $Q_{int} = Q_{FE}$ enter into the first quadrant (third quadrant) the $I_{on}$ ($I_{off}$) will decrease (increase) which is unfavourable to NCFET [17]. Moreover, the smaller AFE which will lead to larger negative slop of S-curve will result in the increase of $V_t$, $I_{on}$, and N-DIBL and the decrease of $I_{off}$, and SS due to greater amplification in $V_{int}$. Therefore, the trade-offs are indispensable for application.

Moreover, due to the existence of NC, some unconventional effects arise in NC VNW-FET electrical properties. Fig.5 (a) shows apparent negative DIBL effect. The $V_t$ of NC VNW-FET increases with the increased drain voltage ($V_{ds}$). Another unconventional effect is negative differential resistance (NDR) effect as shown in Fig.5 (b). The arising of negative DIBL and NDR is due to coupling capacitance from drain to gate [6], [18].

## IV. DEVICE-CIRCUIT CO-ANALYSIS

According to the previous section, the large shift of $V_t$ is occurred for NC VNW-FET comparing to conventional VNW-

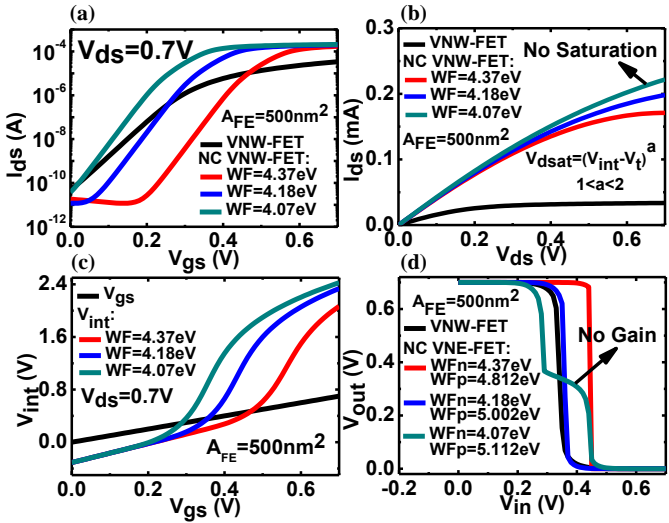

Fig.6 (a) The $I_{ds}$-$V_{gs}$ characteristics of NC VNW-FET for several metal gate WFs with $A_{FE}$ = 500nm$^2$. (b) The $I_{ds}$-$V_{ds}$ characteristics of NC VNW-FET for several metal gate WFs. (c) Voltage in internal gate ($V_{int}$) for different metal gate WFs (d) Voltage transfer characteristics of inverter for NC VNW-FET with different metal gate WFs.

FET. Thus, the $V_t$ of NC VNW-FET need to be adjusted by changing gate WF for low power application. Fig.6 (a) shows the $I_{ds}$-$V_{gs}$ characteristics of NC VNW-FET for several gate WFs at $A_{FE}$ = 500nm$^2$. The $V_t$ decreases with the decreasing of gate WF (for NC nVNW-FET). However, at low gate WF, the $V_{int}$ becomes too large (Fig.6 (c)) to significantly impact saturation voltage of drain ($V_{dsat}$). Generally, the $V_{dsat}$ can be written as $V_{dsat} = (V_{int} - V_t)^a$, where $V_t$ is the threshold voltage of the conventional transistor and a is exponent dependent on mechanism of the current saturation in transistor [19]. If $V_{dsat}>V_{dd}$ (supply voltage) due to the increasing of $V_{int}$ when NC is introduced, the output characteristics exhibit no saturation (Fig.6 (b)). As shown in Fig. 6 (d), the voltage gain is reduced due to the no saturation and then the NC VNW-FET-based inverter is not good binary logic gate. Moreover, the noise margin will be reduced and the leakage power increased if the voltage gain is small. Thus, the co-optimization of $C_{FE}$ and gate WF is crucial to ensure that the inverter transistors can reach saturation in the output characteristics and has appropriate $V_t$.

Fig. 7 (a) shows the $I_{ds}$-$V_{gs}$ characteristics of NC VNW-FET at $A_{FE}$ = 700nm$^2$ and WF = 4.180eV. The $V_t$ of NC VNW-FET is less than conventional VNW-FET (Fig.7 (a)), and the transistor shows saturation in the output characteristics (Fig.7 (b)). The SS of the NC VNW-FET was reduced to 48mV/decade. The voltage transfer curves (VTC) for both NC VNW-FET and conventional VNW-FET inverters at several $V_{gs}$ biases were plotted in Fig.7 (c). It is worth noting that the VTC of inverter exhibits hysteretic at $A_{FE}$=700nm$^2$, even though the corresponding devices are non-hysteretic. The hysteretic circuit response is attributed to NDR. This hysteresis leads to higher noise margins and hence better noise immunity which is similar to a Schmitt trigger [6], [20]-[21]. The maximum voltage gain in NC VNW-FET inverter is about 54, which is much large than the voltage gain, 27, of conventional VNW-FET inverter, as shown in the insert of Fig.7 (d), because

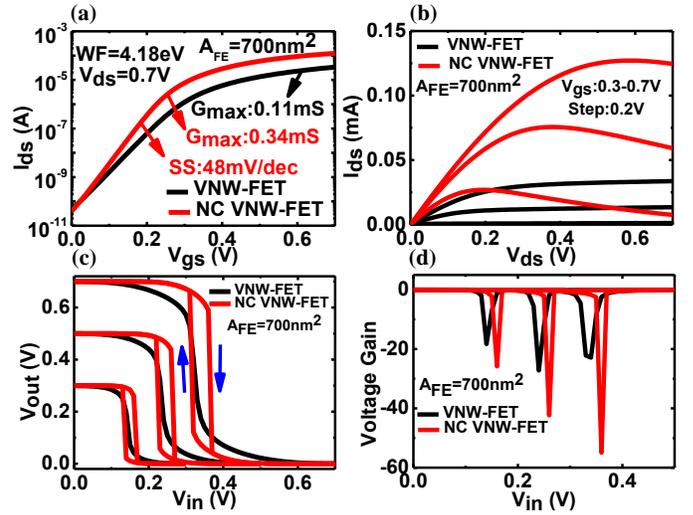

Fig.7 (a) The $I_{ds}$-$V_{gs}$ characteristics of NC VNW-FET at $A_{FE}$ = 700nm$^2$ and WF = 4.180eV. (b) The $I_{ds}$-$V_{ds}$ characteristics. (c) VTC of inverter. (d) Voltage gain.

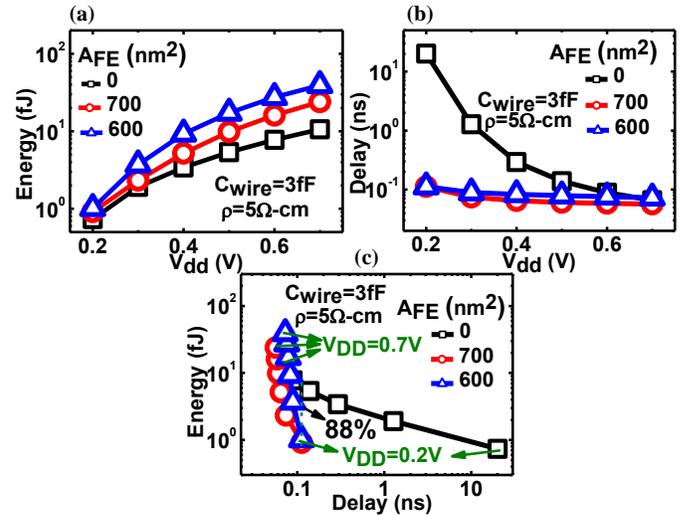

Fig.8 The 7-stage RO characteristics based on conventional VNW-FET and NC VNW-FET. (a) The total energy characteristics. (b) The propagation delay characteristics. (c) The energy-delay characteristics.

of the higher transconductance ($G_{max}$) (Fig.7 (a)) and lower output conductance (Fig.7 (b)) of NC VNW-FET.

A comparison of the characteristics of NC VNW-FET ring oscillator (RO) with conventional VNW-FET RO was made in Fig.8 (a-c). Of note, the capacitance of wire was set to 3fF, and the value of ρ was estimated as mentioned [22]. Comparing to the conventional VNW-FET ROs, the NC VNW-FET ROs exhibit larger dynamic energy owing to higher gate capacitance and lower delay owing to high on current (as shown in Fig.8 (a-b)). It turned out that significant energy reduction, up to 88%, at iso-delay for NC VNW-FETs at low supply voltage is achieved thanks to the on current and SS improvements.

## V. CONCLUSION

In this study, a SPICE model for NC VNW-FETs based on BSIM-CMG model and LK equation was presented, and the model can be applied at sub-3nm node. Suffering from the limitation of gate length of VNW-FETs there is lack of controllable and integrative structures to combine VNW-FETs with ferroelectric material. A new kind of structure was proposed for high performance NC VNW-FETs. The S-curve was divided into four regions and explained from mathematical model and physical mechanism. Based on S-curve four regions theory, a few design rules of NCFETs were proposed. Based on DC analysis of NC VNW-FET inverter, the co-design of gate WF and NC is investigated. A ring oscillator was simulated to analyze the circuit energy-delay, and it shows that signification energy reduction, up to 88%, at iso-delay for NC VNW-FETs can be achieved. This study gives a credible method to analysis the performance of NCFET-based devices and circuits and reveals the potential of NC VNW-FETs in low-power applications.